\documentclass[aps,showpacs,twocolumn,superscriptaddress,nofootinbib,floatfix,section 10pt]{revtex4-2}
\usepackage{amsfonts,amssymb,stmaryrd,latexsym,amsmath,braket}
\usepackage{graphicx,subfigure}
\usepackage{comment}
\usepackage{times}
\usepackage{slashed}
\usepackage{bm}
\usepackage{braket}
\usepackage{appendix}
\usepackage{enumitem}
\usepackage{multirow}
\usepackage{array}
\usepackage{soul}

\usepackage[colorlinks=true,backref=true,linktocpage=true,citecolor=blue,urlcolor=blue,linkcolor=blue,pdfpagemode=UseOutlines]{hyperref}
\usepackage[dvipsnames]{xcolor}

\begin{document}
\title{Corrections to adiabatic behavior for long paths}
\author{Thomas D. Cohen}
\email{cohen@umd.edu}
\affiliation{Department of Physics and Maryland Center for Fundamental Physics, University of Maryland, College Park, MD 20742 USA}

\author{Hyunwoo Oh}
\email{hyunwooh@umd.edu}
\affiliation{Department of Physics and Maryland Center for Fundamental Physics, University of Maryland, College Park, MD 20742 USA}

\begin{abstract}
The cost and the error of the adiabatic theorem for preparing the final eigenstate are discussed in terms of path length. Previous studies in terms of the norm of the Hamiltonian and its derivatives with the spectral gap are limited in their ability to describe the cost of adiabatic state preparation for certain physically large systems. We argue that total time is not a good measure for determining the computational difficulty of adiabatic quantum computation by developing a no-go theorem. From the result of time-periodic Hamiltonian cases, we suggest that there are proxies for computational cost which typically grow as path length increases when the error is kept fixed and small and consider possible conjectures on how general the behavior is.

\end{abstract}

\date{\today}

\maketitle
\section{Introduction}

The adiabatic theorem is nearly as old as quantum mechanics itself, having been proposed by Born and Fock in 1928~\cite{Born1928}, just three years after Schr\"odinger's wave equation~\cite{Schrodinger:1926gei} and Heisenberg's matrix mechanics~\cite{Heisenberg1925}. In recent decades, the adiabatic theorem has become increasingly important as a basis for state preparation on quantum computers
and can serve as a direct method of quantum computation for problems of interest~\cite{farhi2000quantum, doi:10.1137/060648829, doi:10.1126/science.1057726, Childs:2001ge, 959902, 1366223, RevModPhys.90.015002}.

The Hamiltonian never evolves infinitely slowly and diabatic excitation
amplitude (errors) is unavoidable during the traversal of the parameter space of the Hamiltonian. Formal bounds on such errors were first developed decades after the adiabatic theorem itself was proposed~\cite{doi:10.1143/JPSJ.5.435}. Numerous approaches to this were subsequently developed~\cite{10.1063/1.2798382, messiah61,  Wiebe_2012, PhysRevA.80.012106, Burgarth2022oneboundtorulethem};  these relate the error to the total evolving time with the norm of the Hamiltonian and its time derivatives as well as the spectral gap, $\Delta$. Typically such studies find an upper bound on the total evolution time $T$ needed to ensure that a desired maximum error, $\epsilon$, is obtained. 

The utility of these bounds is limited since the norm of the Hamiltonian depends on the eigenvalue of the highest lying states. While systems that are simulated on a quantum computer will, of necessity, have a finite Hilbert space and thus a largest energy eigenvalue, in many circumstances, simulations on quantum computers  
are designed to approximate systems with much larger (or infinite) Hilbert spaces and much higher (or infinite) highest energy states: the closer simulations approximate the physical system of interest, the weaker these formal bounds become. This becomes acute for systems of large or infinite physical extent, such as quantum field theories~\cite{Jordan:2012xnu, Jordan:2011ci, Jordan:2014tma, Liu:2020eoa, Chakraborty:2020uhf, Buser:2020uzs, Gharibyan:2020bab, Kreshchuk:2020dla, Honda:2021aum, Ciavarella:2022qdx, Li:2022ped, Turco:2023rmx, Farrell:2024fit}.


This paper focuses on a different aspect of errors: the interplay among the size of the error, the total computational cost and the path length in parameter space (where the parameter specifies the Hamiltonian). Of course, quantifying the path length is meaningless unless it is specified in a reparameterization-invariant manner. Fortunately, this is possible. The goal is to find a dimensionless proxy for total computational costs instead of the conventional time scale and expect its behavior in the long path length limit.

While the implementation of adiabatic evolution on digital quantum computers has also been considered~\cite{Barends2016, PhysRevLett.116.080503, Wan:2020fwj, Sun_2020, CoelloPerez:2021jkh, Yi:2021fnz, Kovalsky:2022wcy}, a natural context to consider adiabatic evolution is on analog quantum computers. This has the virtue that one can analyze the behavior without the need to analyze the errors associated with Trotterization~\cite{Trotter:1959, Suzuki:1976be}. Of course, one can generalize the analysis here to digital quantum computers by separately considering the computational cost associated with Trotterization errors, but costs considered here apply regardless.

The generic behavior of systems with long path length is important as many physically relevant problems (such as gapped latticized local quantum field theories~\cite{Cohen:2023dll}) have long path lengths. In a sense the goal of this paper is to describe the scaling of the computational cost with path length at fixed small error. One might think that time seems to be a reasonable proxy for computational difficulty of adiabatic quantum computing. However, as will be discussed, this is not really the case. One of the key observations in this paper is the existence of a no-go theorem on scaling with time: there always exists a time-dependent Hamiltonian that violates any proposed bound on asymptotic scaling of time as a function of path length at fixed error. Thus there is no bound on scaling of time with path length that holds generically given a fixed error. Instead, for any given time-dependent Hamiltonian, there exists a time-dependent Hamiltonian that does precisely the same calculation with precisely the same error but does so in a manner in which the time scales with path length in any manner that one chooses. An obvious consequence of this is that if one takes a computational time as the key proxy for computational costs, then no clear relationship between costs and path length can be established. Thus one central purpose of this paper is the identification of a dimensionless quantity that essentially measures how adiabatic the transversal of the path is and proposes this as a proxy for computational cost.    

\subsection{Defining error}


The definition of error depends on which state is being evolved. Since the ground state of the system is often the state of interest, the discussion in this paper focuses on the ground state. However, the formulation goes through for any state {\it mutatis mutandis}.

The error is given as
\begin{equation}
\epsilon \equiv \left \lVert \left (1-|g_f\rangle \langle g_f |  \right) U(t_f,t_i)  |g_i \rangle \right \rVert ,
   \label{Eq:errordef}
\end{equation}
where $|g_f\rangle$ and $|g_i\rangle$ are the ground states of the final and initial Hamiltonians respectively, and $U(t_f, t_i)$ is the time evolution operator; it measures the amplitude for the component of the evolved state that is not in the local ground state.

\subsection{An apparent paradox}

It is worth noting an apparent paradox: intuitively very short paths where relatively little evolution of the state occurs between the initial and final Hamiltonians will have smaller $\epsilon$ for the same $T$ than for a longer path; longer paths have more opportunities for errors to arise. This intuition seems to be in conflict with a formal result that has long been known~\cite{LENARD1959261, GARRIDO1962553,10.1063/1.1704127, Nenciu1981, Nenciu1991, Nenciu1993, HAGEDORN2002235, 10.1063/1.4748968} 
of the asymptotic scaling of errors with time---the so-called switching theorem: there is an asymptotic series\footnote{It gives the error associated with switching a time-dependent Hamiltonian on and off, as one would need to do in adiabatic quantum computing.} for the time-evolved projection operator $U(t_f,t_i) |g_i \rangle \langle g_i |U^\dagger(t_f,t_i)$ which implies that asymptotic series for the operator and error are\footnote{It is not a strict asymptotic series in the sense that coefficients depend on T which only appears in a phase.}
\begin{subequations}
\begin{align}
U(t_f,t_i) |g_i \rangle \langle g_i |U^\dagger(t_f,t_i)& = |g_f \rangle \langle g_f | + \sum_{n=1}^{n_{\rm max}(T)}\frac{B_n}{T^n} + R, \label{Eq:seriesOp}\\
\epsilon  =  \sum_{n=1}^{n_{\rm max}(T)} \frac{b_n}{T^n} + r, & \label{Eq:series}
\end{align}
\end{subequations}
where $|g_i \rangle$ is the initial state, $T\equiv t_f-t_i$, $B_n$ are operators depend on T only with a phase factor, $n_{\rm max}$ is the maximum number of terms for which this asymptotic series improves the description and $R$ is the remainder, an operator that scales with $T$ slower than any power law, while $b_n$ and $r$ are numbers, with $b_n$ dependent of $T$ only in a phase and $r$ representing a remainder that scales slower than any power in $T$. The operators $B_n$ and coefficients $b_n$ only depend on the properties of the trajectory at its end points:  
\begin{equation}
    B_n= B_n^+ - B_n^- , \label{Eq:endpoints}
\end{equation}
where $B_n^+$ depends solely on local properties at the  final time while $B_n^-$ depends on the same properties evaluated at the initial time with  analogous  behavior for the $b_n$\footnote{For example, a straightforward but somewhat involved calculation yields an explicit form for $b_1$: $b_1= \lVert  \sum_{j \neq g} |j_f \rangle (  {\rm e}^{i w_{j,g} T} \frac{1}{\Delta_{j,g}(\lambda_f)}  \langle j | \frac{d}{d\lambda} | g \rangle  |_{\lambda=\lambda_f} - \frac{1}{\Delta_{j,g}(\lambda_i)} \langle j | \frac{d}{d\lambda} | g \rangle  |_{\lambda=\lambda_i} ) \rVert$, where $f$ and $i$ refer to initial and final respectively, $\Delta_{j,g}(\lambda) \equiv E_j(\lambda)-E_g(\lambda)$, and $w_{j,g}  \equiv \int_{\lambda_i}^{\lambda_f} d\lambda \; \Delta_{j,g}(\lambda)$. 
}. 
The fact that the remainder is smaller than any power law in $1/T$ implies that unless all the power-law terms in the series vanish, at sufficiently close to the adiabatic limit of long times the error will always be dominated by the endpoints of trajectory and, thus, errors are independent of the path length.

While this is true, it is misleading if one takes it to mean that the path length is irrelevant. In the first place there are many situations of physical interest where every term in the asymptotic series vanishes and the entire result is contained in $R$. Moreover, if one's interest is the time needed to achieve a fixed small level of error and the time needed for the $R$ term to achieve this diverging as $L \rightarrow \infty$, then a key issue is  the scaling of the coefficients $b_n$ scale with $L$.

\subsection{Defining path length and velocity}

To go further, it is necessary to define a path length. The definition depends on the particular state being evolved (and we focus on the ground state here).  

Consider a family of Hamiltonians, parameterized by a variable $s$, $H(s)$, where $H(s_i)$ is the initial Hamiltonian and $H(s_f)$ is the final Hamiltonian. $H(s)$ is assumed to have a discrete and non-degenerate spectrum for all $s$ between $s_i$ and $s_f$\footnote{The results obtained applies even with weaker conditions but the condition given here simplifies the analysis.} and $H(s)$ is assumed to be infinitely differentiable with respect to $s$. The ground state for the Hamiltonian $H(s)$ is denoted as $|g(s)\rangle$. We are interested in situations where Hamiltonian evolves in time by traversing the path in $s$; it is specified by $s(t)$ from $t_i$ to $t_f$. Note that the relation between $s$ and $t$, $s(t)$, specifies how rapidly the Hamiltonian changes as it traverses the path.

The definition of the length of the path between two points $a$ and $b$ associated with $s_a$ and $s_b$ with $s_a \le s_b$ is given by~\cite{10.5555/2011804.2011811}
\begin{subequations}
\begin{align}
L_{a,b} & \equiv \int_{s_a}^{s_b}  d s \, \lVert   | g'(s) \rangle -| g(s) \rangle \langle g(s)| g'(s) \rangle  \rVert,  \label{Eq:L0} \\
{\rm with} \; & \; | g'(s) \rangle  \equiv \frac{d}{d s} |g(s)\rangle. \label{Eq:g'(s)}
\end{align} 
It is easy to see that the path length is independent of the choice of parameterization: reparameterizing the trajectory in terms of a new parameter that is an invertible and differentiable function of $s$ leaves $L_{a,b}$ unchanged for all $a$ and $b$.  

Note that the fact that $|g(s)\rangle$ is the ground state of $H(s)$ does not fully specify it; its overall phase is not fixed. One can exploit this and choose the phase so that at all points along the path $\langle g(s)| g'(s) \rangle =0$ and the path length becomes
\begin{equation}
L_{a,b}  = \int_{s_a}^{s_b}  d s \, \lVert  | g'(s) \rangle  \rVert.  \label{Eq:L} 
\end{equation}
Of course in practical calculations there is no need to require  $\langle g(s)| g'(s) \rangle =0$, 
but it simplifies the formal analysis and is adopted in the remainder of the paper. One implication of this choice is that in situations where a trajectory passes through the same Hamiltonian more than once, i.e., if $H(s_a) = H(s_b)$ with $s_a \ne s_b$, $|g(s_a) \rangle$, $|g(s_a) \rangle$ and $|g(s_b) \rangle$ might differ by a Berry phase~\cite{Berry:1984jv}.  

We will also choose a phase convention for excited states in analogy to the ground state so that $\langle e_i(s)| e_i'(s) \rangle =0$ where $e_i$ is an arbitrary excited state along the path. It is useful for the purpose of formal analysis to exploit the reparmeterization invariance and 
re-express the trajectory in terms of $\lambda$, defined as 
\begin{equation}
\lambda(s) \equiv \lambda_i + \int_{s_i}^s d \tilde{s} \; \lVert  |g'(\tilde{s})\rangle   \rVert  \; {\rm with} \; L_{a,b}  = \lambda_b-\lambda_a \; ; \label{Eq:lambda}
\end{equation}
$\lambda$ simply parameterizes the length along the trajectory and
the total path length is given by $L \equiv \lambda_f- \lambda_i$. The motion along the trajectory in parameter space is given by $\lambda(t)$.
\end{subequations}

One can introduce a ``velocity'', the rate at which the trajectory in parameter space is traversed.  
The instantaneous velocity at any point along the trajectory, $v(\lambda)$, and the average velocity over the full path, $\overline v$, are given by
\begin{equation} 
    v(\lambda) \equiv \frac{d \lambda}{d t} \; \; {\rm and }\; \; \overline{v} \equiv \frac{L}{T} \;, 
\end{equation}    
where $T \equiv t_f-t_i$ and $L$ is the total path length.
The ``velocity'' has a dimension of inverse time as $L$ is dimensionless. One can factor out $\overline {v}$ as an overall velocity scale, and study the behavior as $\overline v \rightarrow 0$ but following the same trajectory at the same relative rate; thus, the local velocity $v(\lambda)$ will be taken to be proportional to $\overline v$ with a proportionality factor dependent on $\lambda$.   

\subsection{A counterexample for scaling of time in path length}

In the context of quantum computing, 
multiple methods have been proposed for evolving eigenstates of an initial Hamiltonian to a final one, including methods based on the quantum Zeno effect (QZE)~\cite{ PhysRevA.63.052112, Lin2020optimalpolynomial, 10.5555/2011804.2011811} and improvements on these methods based on a form of reflection~\cite{Cohen:2023dll}. For long paths, the reflection-improved-quantum-Zeno (RIQZ) approach of~\cite{Cohen:2023dll} has very efficient scaling: computation resources grow linearly with path length. It is important to see whether the adiabatic approach scales with $L$ in a manner that can match this for long paths.

The significance of the path length in the state preparation along a path of Hamiltonian was emphasized more than a decade ago in a number of insightful papers~\cite{PhysRevA.81.032308, 10.5555/2011804.2011811, boixo2010fast, PhysRevA.89.012314, Bukov:2019fdp, PhysRevA.93.052107}. The strongest claim comes from~\cite{PhysRevA.81.032308}, which uses a clever argument based on computational costs: by exploiting Grover's search algorithm~\cite{Grover1996AFQ}. Ref.~\cite{PhysRevA.81.032308} claims in its abstract to have proved that, ``for constant gap, general quantum processes that approximately prepare the final eigenstate require a minimum time proportional to the ratio of the length of the eigenstate path to the gap.'' Thus, no rigorous adiabatic condition can yield a smaller cost; similar claims exist in multiple places in the paper. However, the validity of this claim has been questioned~\cite{Lychkovskiy2018}. A significant issue is how such a claim should be understood. Superficially the claim might naturally appear to mean that it is impossible to find {\it any} Hamiltonian path (with constant gap) for which the minimum time needed to achieve a fixed level of accuracy in adiabatic state preparation must grow with the path length. However, given the methods employed to make the proof in Ref.~\cite{PhysRevA.81.032308}, such an interpretation of the claim's meaning is not viable. In fact, it is quite easy to construct explicitly systems which violate the putative bound that Ref.~\cite{PhysRevA.81.032308} might appear to  be claiming.

Consider, for example, a 3-level system whose time-dependent Hamiltonian in matrix form is given by 
\begin{widetext}
\begin{equation}
\begin{split}
& H(t) =
\Delta \left( 1 + \frac{t}{\tau} \right) 
\left(
\begin{array}{ccc}
\cos\left (\frac{t}{\tau} + \frac{t^2}{2 \tau^2}  \right) & 0 & \sin\left (\frac{t}{\tau} + \frac{t^2}{2 \tau^2}  \right)\\ 
0 & 1 & 0\\ 
\sin\left (\frac{-t}{\tau} - \frac{t^2}{2 \tau^2}  \right) & 0 & \cos
\left (\frac{t}{\tau} + \frac{t^2}{2 \tau^2}  \right) 
\end{array} \right ) \left (
\begin{array}{ccc}
0 & 0 & 0 \\
0 & \frac{1}{1 + \frac{t}{\tau}}  & 0\\ 
0 & 0 & 2
\end{array} \right ) \left (
\begin{array}{ccc} 
\cos\left (\frac{t}{\tau} + \frac{t^2}{2 \tau^2}  \right) & 0 & \sin\left (\frac{-t}{\tau} - \frac{t^2}{2 \tau^2}  \right) \\ 
0 & 1 & 0\\ 
\sin\left (\frac{t}{\tau} + \frac{t^2}{2 \tau^2}  \right) & 0 & \cos
\left (\frac{t}{\tau} + \frac{t^2}{2 \tau^2}  \right)  
\end{array} \right ) 
\end{split}\label{Eq:3-level}
\end{equation}
\end{widetext}
where $\Delta$ is a constant with dimensions of energy
and $\tau$ is a constant with dimensions of time. The gap in this model is $\Delta$ for all time. If the system is in the ground state at $t=0$, it is straightforward to show that the path length, $L$ (as defined in Eq.~(\ref{Eq:L0})), as a function of time is given by
\begin{subequations}
    \begin{equation}
        L(t) =\frac{t}{\tau} + \frac{t^2}{2 \tau^2},
    \end{equation}
while  $\epsilon$ (defined in Eq.~(\ref{Eq:errordef})) will reach its maximum value 
\begin{equation}
 \epsilon_{\rm max} =\frac{1 }{\sqrt{1 + \Delta^2 \tau^2}}
\end{equation}
during the evolution and it will achieve this value infinite times as $L \rightarrow \infty$; we consider a time when the error is at this value (or close to it).
Taken together this means that given a fixed error and with a constant gap, the time needed to reach $L$ with this error is given by
\begin{equation}
\begin{split}
 t(L) &= \sqrt{2 L \tau^2} \left( \sqrt{1+ (2 L)^{-1}} - \sqrt{(2 L)^{-1}} \right) \\
 &\underset{L \rightarrow \infty}\longrightarrow  \sqrt{2 L \tau^2} .
 \end{split}
\end{equation}
\end{subequations}
Thus this system has the scaling of time with length, at fixed gap $\Delta$ and fixed error, is $L^{\frac12}$ which is slower than linear---and thus contradicts the apparent claim of Ref.~\cite{PhysRevA.81.032308} that the scaling must be at least linear.

Ultimately the simple model above illustrates a central issue. The error that develops when a path of length $L$ is traversed via Hamiltonian evolution depends on both the ``velocity'' of the evolution (as a function of position along the path) and the characteristic energy differences relevant to the non-adiabatic transitions. Indeed, since the time is not dimensionless some property of the system must provide the dimension scale, and the energy differences in the Hamiltonian serve this purpose. Ref.~\cite{PhysRevA.81.032308} implicitly assumes that the scale of these energy differences is fixed by the gap $\Delta$, but as shown in the model above this does not need to be the case. In fact, it is trivial to construct distinct systems whose evolutions are mathematically equivalent by having the local velocity along the traversal and the characteristic energy differences tied to one another in a prescribed manner. By doing this one can engineer any scaling of time as a function of $L$ at fixed error, as will be shown in Sec.~\ref{Sec:No-go}. Moreover, for a wide class of such system the spectral gap is unaffected by such a construction while the relevant energies do scale so that one cannot use the spectral gap to set the scale. A natural conclusion of this is that the scaling of the time with $L$ with fixed error by itself does not provide a useful measure of the computational difficulty of achieving a fixed-error traversal of a path. It is important to come up with a more useful measure of computational difficulty.

\subsection{Plan of the paper}

In the next section, we present a rigorous demonstration that for the special case of paths repeatedly cycling through a Hamiltonian that is periodic in time, the time needed to achieve a fixed small error in nearly adiabatic evolution grows faster than linearly with the path length. This problem is of no practical importance in the context of quantum computing, but it is useful as a benchmark. This situation has the virtue that subtleties associated with identifying a useful proxy of computational difficulty does not arise: with any sensible identification the total cost to traverse $N$ periods with small error will be $N$ times the cost to traverse single period (with the total error proportional to $N$).

Following this is a section which proves a simple theorem: systems in which relevant differences of energy levels along the path are chosen to be proportional to the velocity at each point along the path will lead to errors as a function of path length being independent of the velocity along the path. This implies that one can engineer systems that have scaling behavior of time as a function of $L$ at fixed $\epsilon$.


The paper concludes with a brief discussion of the implications of the analysis for general Hamiltonians.

\section{Time-periodic Hamiltonians}  \label{Sec:TP}

This paper focuses on the regime where $\epsilon$ is small. The time evolution operator is unitary and hence can be written in the form $U(t_f,t_i)=\exp( i A_{f,i})$ where $A_{f,i}$ is hermitian. The small error regime corresponds to situations in which it is sufficient to expand $\exp( i A_{f, i})$ to first order in $A$ in calculating the $\epsilon$ using Eq.~(\ref{Eq:errordef}) and the error becomes 
$\epsilon = \left \lVert \left (1-|g_f\rangle \langle g_f |  \right) A_{f,i}   |g_i \rangle \right \rVert$.  

Suppose that we decompose the evolution operator of a long path into N shorter ones and all the shorter ones are in the regime of small errors:
\begin{subequations}
\begin{align}
& {\rm if} \; \; U(t_f,t_i) = U(t_f,t_{N-1}) U(t_{N-1},t_{N-2}) \cdots  \nonumber\\
& \;\;\;\; \;\;\;\;\;\;\;\;\;\;\;\;\;\;\times U(t_3,t_2)U(t_2,t_1) U(t_1,t_i) \label{Eq:decomp}\\
&{\rm with} \;  U(t_{j+1}, t_j) =\exp \left (i A_{j+1, j} \right), \\
&{\rm then} \;  \epsilon = \left \lVert \left (1-|g_f\rangle \langle g_f |  \right) \sum_{j=0}^{N-1} A_{j+1, j}  |g_i \rangle \right \rVert . \label{Eq:A}
\end{align}
\end{subequations}

Consider the special case where the Hamiltonian is periodic in time with $H(t+\tau) = H(t)$ and examine the trajectory of precisely $N$ periods, beginning with the system in the local ground state. Let $L_{\rm sc}$ be the length of a single cycle so length $L$ is $ L=N L_{\rm sc}$ and assume that the trajectory is nontrivial (where a trivial trajectory has the second half of the trajectory over one period being the time reversal of the first half to ensure a return to the exact ground state after a cycle).
This problem is of limited interest from the prospective of quantum computing, but it is a very useful benchmark for generic behavior.  

Divide the full $N$-cycle trajectory into $N$ parts as in Eq.~(\ref{Eq:decomp}) with each part corresponding to exactly one cycle: $U(t_f,t_i)=  U^N_0$ where $U_0=\exp \left(i A_0 \right )$ is the evolution for a single cycle. From Eq.~(\ref{Eq:A}) if the error is small it is given by 
\begin{equation} 
\epsilon= N \epsilon_{\rm sc} \; {\rm where} \; \epsilon_{\rm sc}=
\left \lVert \left (1-|g_0\rangle \langle g_0 |  \right)  A_0  |g_0 \rangle \right \rVert  ;
\end{equation}
$\epsilon_{\rm sc}$ is the error for a single cycle and $|g_0\rangle$ is the ground state at the beginning and end of each cycle.  

Because the beginning and end of each cycle are the same, every $b_n$ coefficient (which depends only on difference of the initial and final points) in the asymptotic series of Eq.~(\ref{Eq:series}) vanishes for $\epsilon_{\rm sc}$ so $\epsilon_{\rm sc} =r_{\rm sc}$ where $r_{\rm sc}$ is the remainder term and decays in $1/\overline {v}_0 = T_0/L_0$ faster than any power law. Thus, assuming the error is small,
\begin{equation}
\epsilon = N \exp \left(- f( 1/\overline{v}_{\rm sc})  \right ) = \frac{L}{L_{\rm sc}} \exp \left (- f(1/ \overline{v}) \right ) 
\label{Eq:per}
\end{equation} 
the second form exploits the fact that the average velocity over a single cycle is identical to the average over many; $f$ is some monotonic positive-valued function satisfying $\log(x) = o(f(x))$ where $o$ indicates little-o notation implying that $f(x)$ grows faster than logarithmically asymptotically. The third form is equivalent and is included for ease of comparison to case where the trajectory is not periodic. Eq.~(\ref{Eq:per}) implies that 
\begin{equation}
  T =  L \,  f^{-1}\left (\log \left( \frac{L/L_{\rm sc}}{\epsilon} \right)  \right) . 
  \label{Eq:perscal}
\end{equation}  
We do not know the explicit form for $f$, which presumably depends on the details of the system. But, for example, if $f(x)=x$ then $T=L \log\left(\frac{L/L_0}{\epsilon} \right)$, which scales faster than $L$. More generally, Eq.~(\ref{Eq:perscal}) implies that the time must increase faster than linearly if one requires the error to remain fixed.  

Underlying this result is a very simple and intuitive idea: errors accumulate as many cycles are traversed at fixed average velocity so that in order to maintain a fixed error while traversing many cycles the error per cycle needs to be reduced, which requires a slower velocity. This in turn implies that the total time to yield fixed error must grow faster than linearly in $L$.  

Intuitively the simple idea that errors accumulate along the path should hold more generally than for periodic Hamiltonians. Thus one should expect that for fixed errors the computational cost should grow faster than linearly in $L$. However, as seen in the simple three level model introduced in Eq.~(\ref{Eq:3-level}), the time can scale more slowly than $L$. This raises an interesting issue of how to generalize the result seen in the periodic potentials to generic cases while taking into account the physics illustrated by the model Eq.~(\ref{Eq:3-level}) that the time with fixed errors can scale more slowly than linearly in $L$.

\section{A no-go theorem for the relation of time, error and path length } \label{Sec:No-go}

As will be shown in this section, the behavior in the model in Eq.~(\ref{Eq:3-level}) reflects a general no-go theorem: there exists a time-dependent Hamiltonian that violates any proposed bound on asymptotic scaling of time as a function of $L$ at fixed small error. Thus it is impossible to find a generic bound on the scaling of $T(L)$ at fixed $\epsilon$---there is a no-go theorem. This does not mean that the intuition that computational costs should grow faster than $L$ is wrong; rather it implies that time is not a good proxy for computational cost.

To understand the no-go theorem it is useful to start with a trivial observation and then consider a more subtle variation.  The trivial observation is if one simply rescales the overall size of a Hamiltonian everwhere along some path by a constant factor, one can increase the velocity by the same factor while achieving exactly the same error; hence if the overall time scale is decreased  by the same factor as the Hamiltonian is increased, the error stays the same. Of course the goal here is not to determine the value of the overall time scale for a given error, but rather to understand how the timescale varies with $L$. Thus, instead of considering an overall rescaling of the Hamiltonian, it is useful to consider a rescaling of the Hamiltonian  whose value depends on the position along the path.  Doing so and keeping the error fixed, alters the relationship between the length and time to traverse the path; this enables the proof of the no-go theorem.

At a technical level the no-go theorem is easy to prove. One starts with the time-dependent Schr\"odinger equation $H |\psi \rangle = i \frac{d}{d t}|\psi \rangle$ and introduces a unitary transformation, $X$, that expresses $H$ in a locally diagonal basis (i.e., diagonal at whatever point in the traversal the system currently is): 
\begin{equation}
H = X^\dagger H_d X , \label{Eq:diag}
\end{equation}
where $X$ depends on $\lambda$ and only implicitly on time. 
$H_d$ is the diagonal matrix of eigenvalues of the original Hamiltonian at any given point in the traversal. The Schr\"odinger equation for $|\phi \rangle \equiv X |\psi \rangle$ becomes  
\begin{subequations}
    \begin{equation}
\left( H_d + i \dot{X}X^\dagger \right ) |\phi \rangle = i \frac{d}{d t} |\phi \rangle \; , \label{Eq.SEQDia}
\end{equation}
where the dot represents time derivatives. The concept of changing the basis to the fixed eigenbasis is well-known in the context of the adiabatic gauge potential~\cite{Kolodrubetz:2017ofs, Sels:2017daz}.
This can be re-expressed in terms of $\lambda$: 
\begin{equation}
\left(\frac{H_d}{v} + i X' {X}^\dagger \right ) |\phi \rangle = i \frac{d}{d \lambda} |\phi \rangle \; , \label{Eq:SELambda}
\end{equation}
where the prime indicates differentiation with respect to $\lambda$ and $\frac{d}{d t} = v \frac{d}{d \lambda}$ and $v$ is a function of $\lambda$. 
\end{subequations}

Consider two different time-dependent Hamiltonians:  $H_A (\lambda)$ with $\lambda(t) =l_A(t)$ where $l_A$ is a particular time-dependent traversal of the path and $H_B (\lambda)$ with $\lambda(t) =l_B(t)$. Suppose further that 
\begin{equation}
\begin{split}
&H_B(\lambda)=\frac{H_A(\lambda) v_B(\lambda)}{v_A(\lambda)}, \\
 & {\rm where} \; v_{A,B} (\lambda) = \left . \frac{d \, l_{A,B}(t)}{d t} \right|_{t=l_{A,B}^{-1}(\lambda) } .
\end{split} \label{Eq:HAB}
\end{equation}
In that case, if $|\phi_A(\lambda) \rangle$ and $|\phi_B(\lambda) \rangle$ satisfy the same initial condition, and each satisfies Eq.~(\ref{Eq:SELambda}) for their associated Hamiltonians, then $|\phi_A(\lambda) \rangle = |\phi_B(\lambda) \rangle$ for all $\lambda$, which in turn means that the solution of the time-dependent Schr\"odinger equations for the two are equivalent:
$|\psi_A(t_A) \rangle=|\psi_B(t_B) \rangle$ when $t_B= l_B^{-1}\left(l_A(t_A)\right )$. If the two systems follow the path of the same length from $\lambda_i$ to $\lambda_f$ and each starts in the ground state and follows the associated time-dependent Schr\"odinger equation then the states at any given value of $\lambda$ are identical for the two systems at all points on the trajectory. In effect, the two systems are doing exactly the same calculation at different physical rates. Thus the errors computed for the two systems are the same.
However the transit times are different:
\begin{equation} \begin{split}
T_A &= \int_{\lambda_i}^{\lambda_f} d \lambda \frac{1}{v_A} \;   , \;  \; T_B = \int_{\lambda_i}^{\lambda_f} d \lambda \frac{1}{v_B}\\
T_B & =T_A \frac{\overline{v}_A}{\overline{v}_B} \; \; {\rm with} \; \; \overline{v}_{A,B} = \frac{L} {\int_{\lambda_i}^{\lambda_f}  \frac{d \lambda}{v_{A,B}}}.
\end{split}
\end{equation}

An obvious consequence is that by choosing $v_B(\lambda)$ to scale with $\lambda$ rapidly enough whatever the scaling of asymptotic traversal time of system $A$ with $\lambda$, the traversal time of system B can be made to scale with $\lambda$ arbitrarily rapidly while maintaining the same error. For example, for a path starting at $\lambda=0$, if one were to let $v_B(\lambda) = v_A (\lambda) (1 +  \lambda^a) $ and $H_B=H_A (1 +  \lambda^a)$, with $a >1 $, then 
at fixed error, then as $\lambda$ gets large, $\frac{T_B}{T_A} $ asymptotically approaches $\lambda^{-a}$ and by choosing $a$ to be large enough one see that if $T_A$ scales as any power law in $\lambda$, then $T_B$ can be made to scale more slowly with $\lambda$ than fixed power law including linearly. Thus, at fixed error there can be no universal bound limiting the scaling behavior of the time with $\lambda$ to be larger than any proposed scaling behavior.

It is straightforward to generalize the construction here if one wishes to keep the spectral gap constant (as was considered in Ref.~\cite{PhysRevA.81.032308}) or of some asymptotically bounded size; simply increase the Hilbert space of system $B$ to include two parts with a Hamiltonian $H=H_{B,1} \oplus H_{B,2}$, where $H_{B,1}$ is related to system $A$ as above while $H_{B,2}$ is time-independent H with a first excited state that fixes the gap for all values of $\lambda$ along the trajectory. It is always possible to do this if the gap in the spectrum of $H_{B,1}$ starts larger than the gap in $H_{B,2}$ and the velocity in $H_{B,1}$ increases with time monotonically.

\section{Discussion} \label{Sec:Discussion}

In Sec.~\ref{Sec:TP} it was shown that for time-periodic Hamiltonians the time needed to ensure that the error is some fixed small value scales faster in linearly in $L$ asymptotically.

The no-go theorem does not necessarily mean that the computational cost of adiabatic state preparation does not scale with $\lambda$. Rather it can simply reflect the fact that traversal time is not a sensible proxy for computational cost when determining the scaling of cost with $\lambda$. Indeed, it has long been understood~\cite{1366223} that time be itself is not such a proxy: time is a dimensionful quantity and another dimensionful quantity is needed to relate time to the error, while $\epsilon$ and $L$ are dimensionless. The suggestion of Ref.~\cite{1366223} is that a useful quantity to represent the cost is $T \cdot {\rm max} \left \vert H \right \vert$, where $T$ is the transit time and max indicates the maximum value of the trajectory. This identification is problematic in that it suffers from the problem that $\left \vert H \right \vert$ is fixed by the largest eigenvalues of $H$ which may be largely irrelevant in the evolution of the ground state.

This raises the obvious question of whether there exists a useful proxy for computational cost that captures the key physics and, if so, what it is. Let us denote such a quantity $Q_D$. At an intuitive level it seems to make sense that any sensible quantity serving as a measure of computational difficulty, and for which a bound on its scaling with $L$ exists, should at least satisfy the following four conditions:

\begin{enumerate}
    \item The quantity should be such that, as with time, $Q_D$ scales faster than linearly with $L$ for time-periodic potentials. \label{cond1}
    \item The quantity should be dimensionless. This is necessary for the measure of computational difficulty to be independent of the details of the specific hardware used to implement the calculation.  \label{cond2}
    \item Two systems related by Eq.~(\ref{Eq:HAB}) should have the same value of the quantity. As noted previously, the two systems are doing exactly the same calculation simply at different physical rates. Clearly in an abstract sense, the two calculations should be regarded as equally difficult.  \label{cond3}
    \item The quantity should reflect the degree to which the system is in the adiabatic limit. This indicates the fact that requiring the system to be close to the adiabatic limit not only demands significant computational resources but also constitutes the central requirement for suppressing errors.  \label{cond4}
\end{enumerate}

The intuitive picture as to why faster-than-linear scaling with $T$ was natural and simple for the time-periodic case: errors accumulate as many cycles are traversed which implies that maintaining a given error over many cycles requires the error in each cycle to be reduced as the path length increases. This necessitates a slower velocity and faster-than-linear growth in time. That same logic applies for $Q_D$. The no-go theorem implies that this intuitive picture cannot be used to generalize the faster-than-linear scaling of $T$ with $L$ for generic non-periodic cases for scaling with time. However, this does not logically rule out the possibility that the simple intuitive picture might hold more generally for some $Q_D$ than for $T$ and it might imply that faster-than-linear growth of $Q_D$ with $L$ holds more generally.

The main argument of the conjecture is that there is a $Q_D$ such that all the conditions hold for general Hamiltonians: curious readers can refer to Appendix for a detailed discussion of the conjecture and $Q_D$. This conjecture implies that the computational difficulty typically scales faster than linearly with $L$. On the other hand, there exists a method in which the computational cost scales linear in $L$~\cite{Cohen:2023dll}, and thus it should be more efficient when $L$ is very long. Moreover if the variant of the conjectures, such as Eq.~(\ref{Eq:conject2}) of Appendix~\ref{Sec:Generic}, holds and the scaling of computational difficulty is linear in $L$ or worse, the method in Ref.~\cite{Cohen:2023dll} would probably still be preferable for very long paths unless one knew {\it a priori} that the scaling was in fact linear.

Clearly more research is needed. Ideally rigorous mathematical methods could establish Eq.~(\ref{Eq:conject2}) or perhaps some of the other forms of conjecture for some choice or choices of $Q_D$. In the absence of a rigorous proof, numerical studies of various tractable small systems could be used to test the various conjectures. While this could not fully establish any of them, it could provide evidence of their applicability. Numerical studies could in principle rule out the conjecture in Eq.~(\ref{Eq:conject2}) if a counterexample were produced.

\begin{acknowledgments}

This work was supported in part by the U.S. Department of Energy, Office of Nuclear Physics under Award Number(s) DE-SC0021143, and DE-FG02-93ER40762.

\end{acknowledgments}

\bibliography{refs.bib}

\newpage
\appendix
\section{The generic case} \label{Sec:Generic}

It is not immediately clear at this stage how to demonstrate formally whether there exist bounds of the scaling of some $Q_D$ with $L$ that is only violated for a narrow and clearly specifiable class of systems when $\epsilon$ is small and held fixed. Given this, it is useful to consider some plausible conjectures for possible bounds on the scaling that should be explored in detail in the future.  

But before considering various conjectures in detail, it is important to explicitly set up a context by which various conjectures may be compared. The basic structure of a conjecture is that the asymptotic scaling of some measure of computational difficulty 
is bounded to grow with the length of a trajectory, $L$, faster than linearly when the error is held fixed. Such conjectures may also include some additional conditions for the conjecture to hold. However, that basic structure does not fully define a conjecture---one needs to specify how one can change $L$, the length of the trajectory to investigate scaling and how the error is fixed to a given value.

The logic underlying such a structure should be clear. As noted above the basic intuition as to why the time scales faster than linearly---that errors continue to accumulate along the trajectory---would be expected to hold more generally, provided all else was the same, but the no-go theorem showed that in general all else is not the same.  The structure considered here is designed to hold all relevant properties fixed except for the overall scale of the velocity with which the trajectory is traversed; this overall velocity is used to fix the error to a predetermined and small value.

Start with a system with a given Hilbert space with a Hamiltonian that varies according to single parameter. Without loss of generality this parameter can be taken to be $\lambda$, with the initial value $\lambda=0$ occurring at $t=0$ and the trajectory specified by $\lambda(t)$, where $\lambda(t)$ is both continuous and differentiable, with $\lambda'(t) \ge 0$.  This implies that $\lambda(t)$ is invertible with $t(\lambda)$ being the inverse function; the velocity as a function of $\lambda$, $v(\lambda)$, is given by $v(\lambda)= 1/ t'(\lambda)$ and by construction $v(\lambda)$ is continuous and non-negative.   
The dependence of the Hamiltonian on $\lambda$ is defined for arbitrarily large values of $\lambda$ so that the limit of large path length is well-defined. In addition to a specification of $H(\lambda)$, a fully defined trajectory requires a velocity function $v(\lambda)$ and an endpoint $\lambda_f = L$ (recall that by convention, initial point $\lambda_i=0$ and $L=\lambda_f-\lambda_i$). Suppose further that one is interested in a family of trajectories that traverse the same path with the same relative velocities but different absolute velocities; one can denote a member of this family by some reference velocity profile $v_{\rm ref} (\lambda)$ and a scaling factor $s_c$:
\begin{equation}
v(\lambda)= \frac{v_{\rm ref} (\lambda)}{s_c}.
\end{equation}

Properties of the evolution, such as the error, depend on $H(\lambda)$, $v_{\rm ref} (\lambda)$, $s_c$ and $L$. 
Thus any quantity, $Q$, associated with the trajectory can be denoted as $Q(s_c,L;v_{\rm ref}, H)$ so that it is a function of $s_c$ and $L$ and a functional of $v_{\rm ref}$ and $H$. Consider $\epsilon(s_c,L;v_{\rm ref}, H)$, the error to be one such quantity. If one keeps $L$, $v_{\rm ref}$ and $H$, fixed $\epsilon$ becomes a function of $s_c$. This mapping is not necessarily one-to-one as multiple values of $s_c$ might lead to the same value of $\epsilon$. However one can always define 
$s_c(\epsilon,L;v_{\rm ref}, H)$ to be the minimum value of $s_c$ that yields a given value of $\epsilon$ with fixed $L$, $v_{\rm ref}$, and $H(\lambda)$. (And in the regime of interest, one expects that when the error is small, it is associated with a unique value of $s$.) Using this one can always replace $s_c$ and express any quantity of interest $Q$ as a function of $\epsilon$ and $L$, and as a functional of $v_{\rm ref}$ and $H$.

Moreover, one expects $Q(\epsilon, L; v_{\rm ref},H)$ to be proportional to $\epsilon$ for small $\epsilon$---in the limit of small errors, the leading error arises from the transitions from the ground state to excited states (which can occur with either sign) but is independent of the amplitude already in the excited states. Thus
\begin{equation}
\begin{split}
&Q(\epsilon, L; v_{\rm ref},H) = q(L; v_{\rm ref},H) \epsilon + {\mathcal O}(\epsilon^2) \\&  {\rm with}  \;
q(L; v_{\rm ref},H) \equiv \lim_{\epsilon \rightarrow 0} \frac{ Q(\epsilon, L; v_{\rm ref},H)}{\epsilon} .
\end{split} \label{Eq:q}
\end{equation}

The conjectures of interest are ones for which some measure of computational difficulty characterizing the full trajectory, $Q_D$, must scale with $L$ faster than linearly when the error is small and the trajectory is held fixed (including relative velocities).
Thus conjectures are of the form
\begin{equation}
\begin{split}
 &\lim_{L \rightarrow \infty} \frac{\lim_{\epsilon \rightarrow 0 } Q_D(\epsilon, L; v_{\rm ref},H) /  \epsilon}{L} \\
 &=\lim_{L \rightarrow \infty} \frac{q_D(L; v_{\rm ref},H)}{L} = \infty \; ,
\end{split}
\label{Eq:conject}
\end{equation}
where $Q_D$ is a quantity that is taken to be a useful proxy for the computational difficulty of traversing the full trajectory.

Specifying a conjecture involves determining the quantity $Q_D$ to which the conjecture applies and identifying conditions that restrict the applicability of Eq.~(\ref{Eq:conject}).

Once $Q_D$ is determined, various versions of the basic conjecture with varying levels of strength can be formulated that depend on the conditions for which it applies: 
\begin{enumerate}[label=\roman*.]
    \item The conjecture applies with high probability to typical systems that arise in physically relevant setting. However, there exist counterexamples that do not satisfy Eq.~(\ref{Eq:conject}), so they cannot be ruled out a priori. \label{Scenario1}
    \item Given $Q_D$, the conjecture applies to all trajectories for all systems except for cases that amount to a set of zero measure. For this version to make sense, one would need a way to enumerate the various possible reference trajectories and Hamiltonians. The conjecture would then be that, if one randomly selected a Hamiltonian and a reference trajectory, the probability of Eq.~(\ref{Eq:conject}) being not satisfied is zero. However, possible counterexample may exist that do not satisfy Eq.~(\ref{Eq:conject}). \label{Scenario2}
    \item  There exists some clearly definable criterion that is computationally easier to determine whether it is satisfied than determining that Eq.~(\ref{Eq:conject}) is satisfied such that if the criterion is satisfied so is Eq.~(\ref{Eq:conject}). There are a few variants of this possibility. \label{Scenario3}
    \begin{enumerate}
    \item If the criterion is not satisfied, neither is Eq.~(\ref{Eq:conject}).
    \item If the criterion is not satisfied, Eq.~(\ref{Eq:conject}) is sometimes satisfied.
    \item If the criterion is not satisfied, Eq.~(\ref{Eq:conject}) is satisfied with high probability for systems of physical interest. However, there exist counterexamples that do not satisfy Eq.~(\ref{Eq:conject}).
    \item If the criterion is not satisfied, Eq.~(\ref{Eq:conject}) is satisfied except for cases that amount to a set of zero measure. This variation requires a way to enumerate the various possible reference trajectories and Hamiltonians.
    \end{enumerate}
\end{enumerate}

One might wish to consider a variant of the conjecture in which Eq.~(\ref{Eq:conject}) holds in all cases a sensible $Q_D$ is identified might exist, but this can probably be ruled out from the beginning. In cases where time-dependent systems when written in the form of Eq.~(\ref{Eq.SEQDia}) has $ H_d + i \dot{X}X^\dagger$ independent of time (a situation that can occur), sensible definition for $Q_D$ (as will be considered in the next section) will have $Q_D$ linear with $L$ at fixed small error but not faster than linear. Thus, the strongest conjectures are likely to be some variant of Scenario~\ref{Scenario3}. However one can construct a separate conjecture that could hold universally. The growth of $Q_D$ with $L$ is linear or faster when the error is small and fixed for all systems and all trajectories
\begin{equation}
    \lim_{L \rightarrow \infty} \frac{\lim_{\epsilon \rightarrow 0 } Q_D(\epsilon, L; v_{\rm ref},H) /  \epsilon}{L} < \infty \; .
\label{Eq:conject2}
\end{equation}
In one sense this is a stronger conjecture than the previous ones because it holds universally, but in another sense it is weaker because it only requires linear or faster scaling rather than faster than linear scaling.

As will be discussed in the next section it is possible to find quantities for $Q_D$ that satisfy all four conditions \ref{cond1}-\ref{cond4}. Moreover, it seems highly plausible that at least some variant of the conjectures should hold for these.

\section{Quantifying adiabaticity}

Note that although total time has been shown not to be viable as a measure of difficulty, part of the reason it might have seemed like a sensible measure is that it is connected to adiabaticity---when all else is equal, longer times correspond to more adiabatic evolution. The problem, however, as exposed by the no-go theorem is that ``all else'' need not be equal. Thus, a key issue is to find a quantity that reflects adiabaticity more accurately than time and does so in a manner consistent with conditions \ref{cond1}-\ref{cond4}.

To do this it is useful to see how adiabatic evolution works. Consider Eq.~(\ref{Eq:SELambda}), whose solution in  the limit of small errors yields an error 
\begin{widetext}
\begin{equation}
\epsilon =\left \Vert \sum_{k \neq g} \int_{\lambda_i}^{\lambda_f} d\lambda \exp\left (i \int_{\lambda_i}^\lambda d\tilde{\lambda} \frac{E_k(\tilde{\lambda})-E_g(\tilde{\lambda})}{v(\tilde{\lambda})} \right )    \langle k(\lambda) | X' {X}^\dagger |g (\lambda) \rangle \right \Vert , \label{Eq:osc}
\end{equation}
\end{widetext}
where the sum over ${k \neq g}$ indicates a sum over excited states. It is clear that the adiabatic behavior occurs---the error becomes small---because the rapid phase oscillations due to energy differences between the ground state and excited states cause large cancellations on the scale at which the Hamiltonian evolves. 

It seems natural to equate the amount of adiabaticity with computational difficulty. Consider for a moment the resources needed for calculations using a digital quantum computer in which the evolution are approximated via discrete time steps. Clearly to accurately reflect nearly adiabatic time evolution one would need to be able to resolve the rapid oscillations of the continuous evolution. Thus the more adiabatic the time evolution is, the larger the number of time steps is needed and so the more computational resources are needed.

At a given $\lambda$, $a_k(\lambda) \equiv (E_k(\lambda)-E_g(\lambda))/{v(\lambda)}$ quantifies how transitions from the ground state to the $k^{\rm th}$ excited state are adiabatically suppressed in the vicinity of $\lambda$.  To quantify how transitions out of the ground states are adiabatically suppressed overall when the system is at $\lambda$, an appropriate weighted average of the   $a_k(\lambda)$ should be taken. A natural weighted average is one in which $w_k(\lambda)$, the weight of the $k^{\rm th}$ state,  is the probability that the operator  $i X' {X}^\dagger$, the term driving the possible transition, takes the ground state to the $k^{\rm th}$ level. With the definition of $\lambda$ in Eq.~(\ref{Eq:lambda}),  $\sum_k | \langle k(\lambda) | g'(\lambda) \rangle|^2=1$. This implies that $w_k=|\langle k(\lambda) | g'(\lambda) \rangle|^2$ is the probability that $i X' {X}^\dagger$ drives a transition to level $k$. Accordingly, 
\begin{equation}
    a(\lambda) = \langle g'| \frac{H(\lambda) -E_g(\lambda)}{v(\lambda)} | g' \rangle
    \label{Eq:al}
\end{equation}
seems to be a reasonable measure for how adiabatic at given value of $\lambda$.

In the present context it is essential to define an overall measure of adiabaticity of a given traversal of the path.  If one denotes this as $Q_D$ then a natural choice  is the integral of   $a(\lambda)$ over the entire path:
\begin{equation}
 Q_D \equiv \int_{\lambda_i}^{\lambda_f} d \lambda  \, \langle g'| \frac{H(\lambda) -E_g(\lambda)}{v(\lambda)} |g' \rangle . \label{Eq:QD1}
\end{equation}

Given that $E_k(\lambda) -E_g(\lambda) \ge \Delta(\lambda)$ where $\Delta (\lambda) $ is the spectral gap at $\lambda$ and $k$ is an excited state, and that $\Delta (\lambda) \ge \Delta$ where $\Delta$ is the minimum spectral gap over the trajectory, it follows that
\begin{equation}
Q_D \ge \int_{\lambda_i}^{\lambda_f} d \lambda  \, \langle g'| \frac{\Delta(\lambda)}{v(\lambda)} |g' \rangle \ge \Delta T \;,
\label{Eq:ineq}\end{equation} 
where $T=t_f -t_i$.
While there is a no-go theorem for the scaling of $T$ with $L$, the direction of inequality 
is such that it does not immediately constrain the scaling of $Q_D$; given what has been shown up to this stage, it remains possible that a bound on the asymptotic scaling of $Q_D$ with $L$ exists consistent with one of the variants.

It is easy to see that $Q_D$ defined in this way satisfies Condition~\ref{cond1} in the previous section: asymptotically $Q_D$ scales faster than linearly with time for time-periodic Hamiltonians.
For cases in which the trajectory covers an integer number of periods, $N_p$, the total time $T=N_p \tau$ where $\tau$ is the period, while $Q_D=N_p {Q_D}_{\rm sc}$ where ${Q_D}_{\rm sc}$ is the value of $Q_D$ over a single cycle. Thus $Q_D=( {Q_D}_{\rm sc}/\tau ) T$ and $Q_D$ is directly proportional to $T$. If the trajectory does not precisely cover an integer number of periods, the proportionality is not exact for finite path lengths, but in the limit of long path lengths, the ratio of $Q_D$ to $T$ asymptotically approaches $( {Q_D}_{\rm sc}/\tau )$ and the proportionality is restored. Since at fixed error $T$ scales faster than linearly in $L$, so does $Q_D$.

$Q_D$ is clearly dimensionless so it automatically satisfies Condition~\ref{cond2}. Moreover, by construction $Q_D$ satisfies Condition~\ref{cond3} and the motivation in picking $Q_D$ was to reflect adiabaticity so Condition~\ref{cond4} is satisfied. This choice for $Q_D$ satisfies all four conditions and thus seems to be a reasonable measure for the computational difficulty.  

It also seems highly plausible that $Q_D$ will satisfy at least one of the faster-than-linear Scenarios spelled out in Sec.~\ref{Sec:Generic}.  The logic underlying this has been mentioned before: the longer the path is in $L$ the more opportunities there are for errors to accumulate.

It should be noted that $Q_D$ as specified in Eq.~(\ref{Eq:QD1}) is not a unique choice for a measure of computational difficulty associated with adiabaticity. For example,
\begin{subequations}
    \begin{align}
  Q_D &\equiv \int_{\lambda_i}^{\lambda_f} d \lambda  \, \frac{\sqrt{\langle g'| \left(H(\lambda) -E_g(\lambda)\right)^2 |g' \rangle }} {v(\lambda)}\label{Eq:QD2}\; \; {\rm and}\\
    Q_D &\equiv \int_{\lambda_i}^{\lambda_f} d \lambda  \, \frac{\left ( \langle g'| \sqrt {H(\lambda) -E_g(\lambda) } |g' \rangle \right)^2} {v(\lambda)}\label{Eq:QD3}
    \end{align}
each could serve sensibly in this role with $Q_D$ as given Eq.~(\ref{Eq:QD2}) emphasizing the contributions of higher energy states than Eq.~(\ref{Eq:QD1}) while Eq.~(\ref{Eq:QD3}) emphasizes the contributions of lower lying states. Both of these choices satisfy Conditions \ref{cond1}-\ref{cond4} of the previous section.
More generally for any  monotonically increasing and positive (and hence invertible) function of positive inputs, $f(x)$,
\begin{equation}
Q_D \equiv \int_{\lambda_i}^{\lambda_f} d \lambda  \, \frac{f^{-1} \left(\langle g'| f\left(H(\lambda) -E_g(\lambda)\right) |g' \rangle \right)} {v(\lambda)} \label{Eq:QD4}
\end{equation}
satisfies  Conditions \ref{cond1}-\ref{cond4} and might reasonably 
\end{subequations}
could serve as $Q_D$. 

It is unclear whether all of these possible choices for $Q_D$ lead to the same behavior.  It is plausible that all of them typically lead to a faster than linear scaling with  $L$ at fixed small error but some choices might lead to more or less restrictive conditions on applicability.

It seems highly plausible that Scenario~\ref{Scenario1}, the weakest form of conjecture in Sec.~\ref{Sec:Generic}, should hold for the various choices of $Q_D$ defined in previous section. If this is correct and if $Q_D$ is really a reflection of computational difficulty, there are potentially serious implications for the advisability of using adiabatic state preparation in situations where the path length is very long. 



\end{document}